 \documentclass[11pt,a4paper]{article}
 \usepackage{amsmath}
\usepackage{cite}
\usepackage{graphicx}
\usepackage{amsfonts}
\usepackage{amssymb}
\input epsf
\begin{document}
\vphantom{0}
\vskip1.1truein
\renewcommand{\theequation}{\thesection.\arabic{equation}}
\newcommand{\eqal}[2]{\begin{eqnarray} #2 \end{eqnarray}}
\newcommand{\eqarr}[2]{\begin{array} #2 \end{array}}
\newcommand{\eqn}[2]{\begin{equation} #2 \end{equation}}
\newcommand{\newsec}[1]{\setcounter{equation}{0} \section{#1}}
\newcommand{\no}{\nonumber}
\newcommand{\tr}{{\rm tr~}}
\newcommand{\Tr}{{\rm Tr~}}

\def\yboxit#1#2{\vbox{\hrule height #1 \hbox{\vrule width #1
\vbox{#2}\vrule width #1 }\hrule height #1 }}
\def\fillbox#1{\hbox to #1{\vbox to #1{\vfil}\hfil}}
\newcommand{\ybox}{{\lower 1.3pt
\yboxit{0.4pt}{\fillbox{8pt}}\hskip-0.2pt}}
\makeatletter
\ifcase\@ptsize
 \font\tenmsy=msbm10
 \font\sevenmsy=msbm7
 \font\fivemsy=msbm5
\or
 \font\tenmsy=msbm10 scaled \magstephalf
 \font\sevenmsy=msbm8
 \font\fivemsy=msbm6
\or
 \font\tenmsy=msbm10 scaled \magstep1
 \font\sevenmsy=msbm8
 \font\fivemsy=msbm6
\fi
\newfam\msyfam
\textfont\msyfam=\tenmsy  \scriptfont\msyfam=\sevenmsy
  \scriptscriptfont\msyfam=\fivemsy
\def\Bbb{\ifmmode\let\next\Bbb@\else
 \def\next{\errmessage{Use \string\Bbb\space only in math 
mode}}\fi\next}
\def\Bbb@#1{{\Bbb@@{#1}}}
\def\Bbb@@#1{\fam\msyfam#1}
\newcommand{\IZ}{{\Bbb{Z}}}
\makeatother

%

%

\def\l{\lambda}
 \def\d{\delta}
 \def\dbar{{\overline\partial}}
 \def\inv{^{-1}}
 \def\hf{\frac{1}{ 2}}
 \def\uq{style{1\over 4}}
  \def\frac#1#2{{\textstyle{#1\over#2}}}
 \def\inv{^{\raise.15ex\hbox{${\scriptscriptstyle -}$}\kern-.05em 1}}

 \def\({\left(}
 \def\){\right)}
 \def\<{\left\langle\,}
 \def\>{\, \right\rangle}
 \def\[{\left[}
 \def\]{\right]}

 \def\Re{{\rm Re ~}}
 \def\Im{{\rm Im ~}}
 
 \def\p{\partial}
 \newcommand\encadremath[1]{\vbox{\hrule\hbox{\vrule\kern8pt
\vbox{\kern8pt \hbox{$\displaystyle #1$}\kern8pt}
\kern8pt\vrule}\hrule}}

 \def\a{\alpha}
 \def\b{\beta}
 \def\g{\gamma}
 \def\d{\delta}
 \def\e{\epsilon}
 \def\eps{\varepsilon}
 \def\th{\theta}
 \def\vt{\vartheta}
 \def\k{\kappa}
 \def\l{\lambda}
 \def\m{\mu}
 \def\n{\nu}
 \def\x{\xi}
 \def\r{\rho}
 \def\vr{\varrho}
 \def\s{\sigma}
 \def\t{\tau}
 \def\th{\theta}
 \def\z{\zeta }
 \def\vp{\varphi}
 \def\G{\Gamma}
 \def\D{\Delta}
 \def\T{\Theta}
 \def\X{\Xi}
 \def\P{\Pi}
 \def\S{\Sigma}
 \def\L{\Lambda}
 \def\O{\Omega}
 \def\oo{\hat \omega   }
 \def\ov{\over}
 \def\o{\omega }
 \def\bbox{{\sqcap \ \ \ \  \sqcup}}
 \def\tria{$\triangleright $}
 \def\sn{{\rm sn} }
   \def\cn{{\rm cn} }
   \def\dn{{\rm dn} }
   \def\z{\zeta }
   \def\uy{u_{\infty} }
   \def\hb{\hbar\ }
   \def\bh{{1\over \hbar}}
    \def\Im{{\rm Im}}
    \def\Re{{\rm Re} } 
   \def\CA{{\cal A}}
   \def\CF{{\cal F}}
   \def\CG{{\cal G}}
   \def\CL{{\cal L}}
   \def\CN{{\cal N}}
   \def\CP{{\cal P}}

\def\IR{ \mathbb{R} }
 \def\gst{\gamma _{\rm str}}
 
\def\gs{g_{_{\rm s}}}
\def\gym{ g_{_{\rm YM}}}
  \def\mb{ {\mu^{_{\ninepoint B}} } }
 \def\mbt{\tilde \mu_{_{\ninepoint B}}}
  \def\nd{ {\rm nd}}
  \def\cs{ {\rm cs}}
  \def\sc{ {\rm sc}}
   \def\nc{ {\rm nc}}
    \def\dc{ {\rm dc}}
      \def\cd{ {\rm cd}}
       \def\sd{ {\rm sd} }

    \def\ds{ {\rm ds} }
\def\IC{\mathbb{C}   }
  \def\IN{\mathbb{N} }
  \def\IP{ \mathbb{P} }
\def\hf{ \frac{1}{2}}
 \def\yy{ Y_{\o } }
        \def\bb{\noindent $\bullet$ \ }
 \def\hx{ \hat x } 
 
\renewcommand{\theequation}{\thesection.\arabic{equation}}
\newcommand{\Det}{{\rm Det~}}


\vskip -2cm

\rightline{SPhT-T06/012}
 
 
\vskip 2cm

\centerline{\Large{\bf  Thermal  flow   in the gravitational $O(n)$ model }}
 
\vskip 2cm
\centerline{Ivan K. 
Kostov\footnote{\texttt{Ivan.Kostov@cea.fr}}\footnote{Associate 
member of {INRNE -- BAS}, \ 
Sofia, Bulgaria } 
}

\vskip 16pt
\centerline{\sl Service de Physique Th{\'e}orique,  CNRS -- URA 2306}
\centerline{\sl  C.E.A. - Saclay, }
\centerline{\sl  F-91191 Gif-sur-Yvette, France}

\vskip 2cm
 
\begin{abstract}
 We study the massless flow from  the critical point (dilute loops) to  the 
low-temperature phase (dense loops) of  the $O(n)$ loop gas model 
when the model is coupled to 2D gravity. The flow is generated by the 
gravitationally dressed  thermal operator $\Phi_{1,3}$ coupled to the 
  renormalized loop tension $\l\sim T-T_c$. We find that the susceptibility
  as a function of the thermal coupling $\l$ and the cosmological 
  constant $\mu$ satisfies a simple transcendental equation.
 \end{abstract}

 \vskip 2cm

 {\small \it Talk delivered at the Fourth 
  International Symposium "Quantum Theory and Symmetries",
  
 Varna, Bulgaria, 15-21 August 2005
 }

  \newpage

\section{Introduction and summary}

 \noindent 
It is well known that the critical phenomena on flat and fluctuating
lattices are deeply related.  For each critical point described by a
``matter'' CFT, the ``coupling to gravity" consists in adding a
Liouville and ghost sectors and dressing the scaling operators by
exponents of the Liouville field.  In such simplest theories of 2D
gravity are quite well understood nowadays due to the progress towards
the exact solution of Liouville theory achieved in the last decade.
  
On the other hand, almost nothing is known about theories of 2D
gravity in which the matter field has massless excitations but is not
conformal invariant.  A typical example of such a theory is the $O(n)$
loop model on a honeycomb lattice \cite{Nien}.  This model has two
non-trivial critical points, the dense and the dilute phases of the
loop gas, described by two different CFT's.  The massless flow
relating these two CFT's is generated by the thermal operator
$\Phi_{1,3}$.  Other theories with massless flows have been studied in
\cite{AlZa, AlZb, FSZ}.
 
The thermal flow in the gravitational $O(n)$ model is expected to be
described by Liouville and matter CFT's coupled through the operator
gravitationally dressed thermal operator $\Phi_{1,3} $.  It is not known
at present how to solve such a theory.  On the other hand, the
corresponding microscopic theory, the $O(n)$ model on random planar
graphs, can be solved exactly using its dual formulation as a matrix
model \cite{KostovFY, GaudinVX, KostovPN, EynardCN}.  The exact
solution in the case of general potential has been formulated in
\cite{ EynardNV, EynardZV}.
   
    The aim of this paper is to work out, using the matrix model formulation, 
  explicit expressions for the partition functions of the gravitational $O(n)$ 
  model on the  disk and on the sphere along the thermal flow.
   In this section we give a brief introduction in the $O(n)$ model and
    present our results.   
       
     \bigskip

 The $O(n)$ model can be defined on any trivalent graph $\CG$.
  The local fluctuating variable associated with the sites $r\in\CG$
 is an $O(n)$ spin with components $S_1(r), \dots S_n(r)$,
normalised so that $\Tr S_a(r)S_b(r')=\delta_{ab} \delta_{rr'}.$ The
partition function on the graph $\CG$  is defined as
\eqal\actOn{
Z_{_{O(n)}}(\CG; T)
=\Tr\prod_{ <rr'> }\big(1+\frac{1}{T}  \sum_{a}S_a(r)S_a(r')\big)
}
where $ T$ is the temperature and the product runs over all links
$<\!\!  rr'\!\!  >$ of  the graph $\CG$.  Expanding the trace as a sum of
monomials, the partition function can be written as a sum over all
configurations of self-avoiding, mutually avoiding loops that can be
drawn on $\CG$,
\eqn\acton{
Z _{_{O(n)}}(\CG, T) 
=\sum _{\rm loops}  T^{-\CL_{\rm tot}}  \, n^{\CN_{\rm loops}},
\label{acton}}
 as shown in Fig.1. Here $\CL_{\rm tot}$ is the total length 
 of the loops, equal to the 
number of occupied lattice edges and $\CN_{\rm loops}$ is the number
of loops.  Unlike the original formulation, the loop gas
representation (\ref{acton}) makes sense also for non-integer $n$
and  has a continuous transition in two dimensions for $|n|\leq2$. 
 In this interval the number of flavors can be parametrized as

%

  \epsfxsize=200pt
 \vskip 20pt
 \hskip 100pt
 \epsfbox{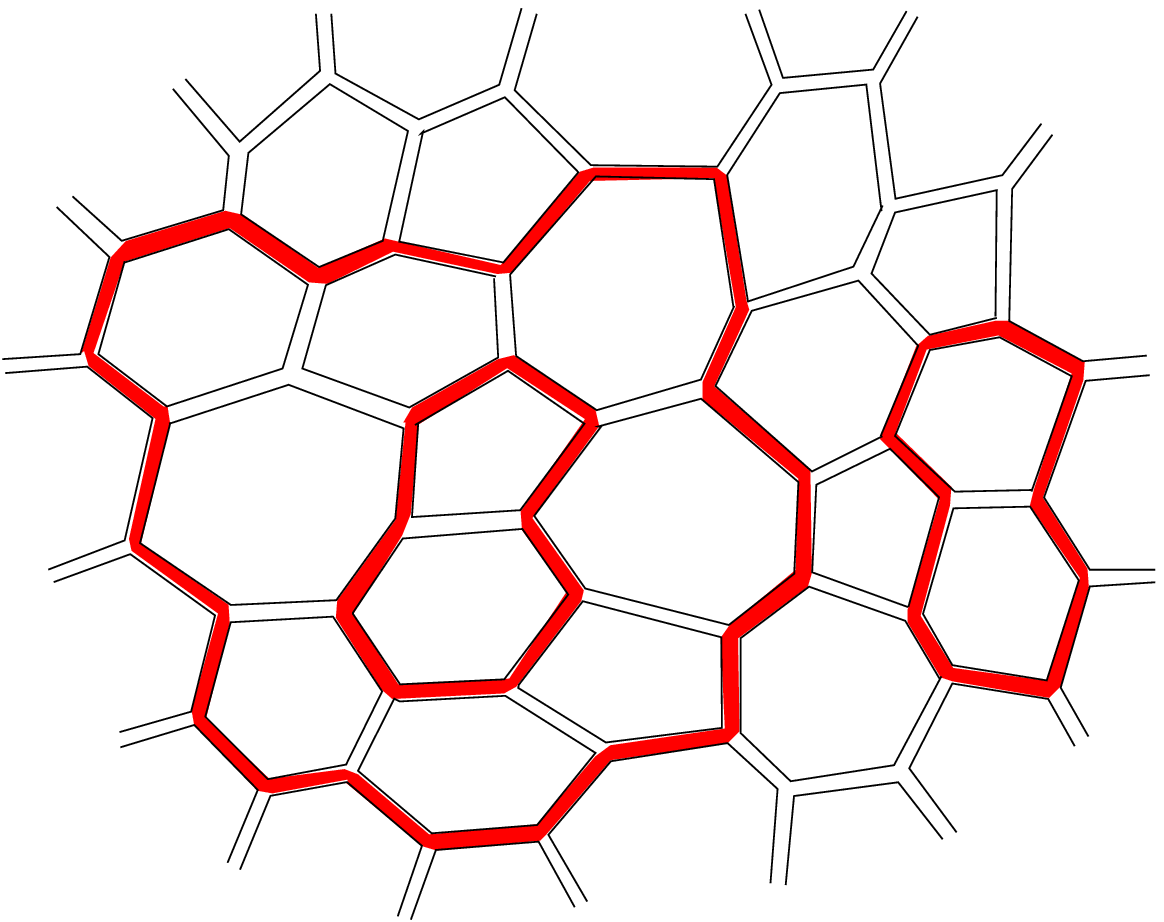    }
 \vskip 5pt
 
 \centerline{Fig. 1:   Loops on a  trivalent fat planar graph     }
  
  \vskip  10pt

  \eqn\defen{
 n = 2 \cos \pi\nu, \ \ \ \ 0<\nu <1.
 \label{nnu}
 } 
The phase diagram of the loop gas on the infinite regular trivalent graph, the honeycomb lattice,  was first established
in \cite{Nien}.  At the critical temperature $T_c = 2\cos
\frac{\pi}{4}\nu$ the loop gas model is solvable and is described by a
CFT with central charge
\eqn\ccrit{
c_{_{\rm critical }}= 1- 6 \frac{\nu^2}{ 1+\nu}.
\label{ccrit}}
For $T>T_c$ the theory has a mass gap.  The low-temperature, or
``dense", phase $T<T_c$  is a flow to an
attractive fixed point \cite{BNh} at
  $T^{\rm dense}_c = 2\sin \frac{\pi}{4} \nu,$ where the theory is
  again solvable and is described by a CFT with smaller central charge
 \eqn\cden{
 c_{_{\rm dense}}= 1- 6 \frac{\nu^2}{ 1-\nu}.
\label{cden}
 }

 The scaling behaviour of the model in the vicinity of a critical
 point is described by an action of the form \cite{FSZ}
  \eqn\aaca{\CA =\CA_{\rm
 critical} + \delta T \int \Phi_{1,3}
\label{aaca}
}
where $\delta T = T-T_c $ and $ \Phi_{1,3}$ is the thermal operator 
with  conformal dimensions  
  \eqn\dtemp{
  \Delta_{1,3}= \bar \Delta_{1,3} = \frac{1-\nu}{1+\nu}.
  }
Microscopically the thermal operator $\Phi_{1,3}$ counts the total
length of the loops $\CL_{\rm tot}$.  Added to the action, it
generates a mass of the loops.  For $ \delta T >0$ the deformation
(\ref{aaca}) describes, going from short to the long distance scales,
the flow to a massive theory with mass gap
\eqn\mssc{ m \sim \delta T   ^{1/(2-2\Delta_{1,3})}= \delta T  ^{1+\nu\over 4\nu} .}
When $\delta T  <0$ the deformation (\ref{aaca}) describes a massless flow
between two different CFT with central charges (\ref{ccrit}) and
(\ref{cden}).  In the CFT for the dense phase, describing the IR
limit, the flow to the attractive critical point is generated by
another, irrelevant, operator, $\Phi_{3,1}$.

   \bigskip
   
   The  $O(n)$ model on a fluctuating  lattice \cite{KostovFY}
can be formulated as a statistical 
ensemble of  trivalent planar graphs covered by self-avoiding and
mutually avoiding loops. The critical thermodynamics  
is controlled by the temperatore $T$ and one extra parameter, the 
 bare  cosmological constant $\k$ coupled to the size of the graph.
 For example, the partition function on the sphere is defined as 
 \eqal\CFF{\CF(\k, T) = 
 \sum_{\CG}    \k^{\CA}\ Z _{_{O(n)}}(\CG, T) 
 \label{lopgs}
 }
where the summation is taken over all connected fat graphs $\CG$
with the topology of a sphere and $\CA$ is the area 
(the number of vertices) of the graph.
          
  As in the case of a flat lattice, the model has
three critical points.  Each critical point is characterized with the
``string susceptibility'' exponent $\gst$ related to the matter
central charge by $ c _{\rm matter} =1-6 \gst ^{2} /(1-\gst).  $ \cite{KPZ}.
Qualitatively, at  the critical point $\k=\k^*, T=T^* $ both the area 
of the graph and the  length of the loops diverge. The flow to the massive, 
high-temperature, phase is along the critical line $\k= \k_{I} (T)$ 
where the area of the graph diverges, while the loops remain finite.  
The flow to the dense,  low-temperature,  phase  is along another line, 
$\k= \k_{II} (T)$,  where the  area of the  graph diverges because of the 
diverging length of the densely packed  loops.  Thus the  continuum limit
is described by the  vicinity of the critical line $\k= \k_c(T)$, which  
consists of two branches meeting at the critical point:
$$\k=\k_c(T) = \begin{cases} \k_{I} (T), & T>T^*;\cr
 \k_{II} (T), & T<T^*.
 \end{cases}
 $$
 The exponent $\gst$
is given in the three phases by
 \eqn\gstde{
\gst = \begin{cases}
    - \hf   & \text{\small\  high-temperature phase (massive loops)};   \\
     -\nu  & \text{\small  \ critical point (dilute loops)}; \\
      -\frac{\nu}{ 1-\nu} &  \text{ \small  low-temperature phase (dense loops).} 
\end{cases}
 }
 In the continuum limit  the loop gas model on a fluctuating lattice is described 
 in terms of the renormalized coupling constants
  \eqal\mulam{ 
 \mu \sim \k_{c}(T) -\k^2,
  \qquad
 \l &\sim 
  T^*-T.
 }
 Denote by $\Phi(\mu, \l, \mu_B)$  the partition function on the disk with boundary 
 cosmological constant $\mu_B$. It is  given by the  Laplace transform of the 
 disk partition function $\tilde \Phi(\mu, \l, \ell) $  with fixed boundary  length $\ell$:
 \eqal\diskpf{\Phi(\mu, \l, \mu_B)=
 \int _0^\infty d\ell \, e^{- \mu_B \ell} \, \tilde  \Phi(\mu, \l, \ell) .
 }
 
 An important characteristics of the model is the  boundary entropy $-M$,
 defined as the exponent in the exponential decay of the 
 disk partition function when the length tends to infinity:
 \eqal\defM{
 M =- \lim_{\ell\to\infty}{ \log \tilde \Phi( \mu, \l, \ell)\over \ell} .
 }
As each loop can be considered as two boundaries glued together, the effective loop 
tension is equal to twise the boundary entropy.

\bigskip

\noindent
 {\it Summary of the results:}

\smallskip

\noindent
We  have found that the susceptibility $ \chi \sim - \p_\mu^2 \CF$
is related to the boundary entropy as
  \eqn\defchi{
  \chi =  M^{2\nu}.
 \label{defchi}
  }
The function $M=M(\l,\mu)$ is determined by comparing the expressions
for the  derivative  $\p_\mu \Phi|_{\mu_B}$  and $\p_{\mu_B}\Phi_\mu$,
which we obtained  by solving  the saddle point equations for the $O(n)$ 
matrix model. The derivatives of the disk partition function  are 
given in parametric form  by
 %
 \eqal\xoft{
 \mu_B&=& M\cosh \t,
\cr
& &
\cr
     \p_{\mu_B} \Phi|_\mu  &=& - 2 M^{1+\nu}     \cosh(1+\nu) \t
\cr
 &&   
 \cr  &&+ 2
\l\,   M^{1-\nu} \cosh (1-\nu)\t    ,   
\cr 
       & &
       \cr
    \p_\mu\Phi |_{\mu_B}
      &=&  \nu^{-1} M^\nu 
         \cosh\nu \t.
         \label{xoft}  
   }
   
   \noindent
  (The normalization of the coupling constants is chosen for our convenience.)
  The compatibility of  these two 
   expressions  implies  the following   transcendental  equation
   for  the boundary entropy:
\eqn\bentr{
\mu = (1+\nu) \, M^2 + \l \, M^{2-2\nu}.
\label{bentr}
}
Equation (\ref{bentr}) generalizes the previously 
obtained results for the scaling  behavior of the boundary entropy 
 in the dilute and the dense phases, 
 $M\sim \mu^{1\over 2}$ in the dilute phase and $M \sim \l^{1\over 2-2\nu}$
in the dense phase \cite{KostovCG}.

As the susceptibility is related to the boundary entropy by (\ref{defchi}),
  it  satisfies the  transcendental  equation
        \eqal\STREQ{
       \mu =(1+\nu)  \chi^{1\over \nu}  +\l  \chi ^{ 1-\nu \over \nu }   ,
  \label{STREQ}
  }
 which coincides, after a redefinition of the 
 variables, with   the eqation 
found in \cite{KKK} for the susceptibility of the gravitational sine-Gordon model.

The dimension of the coupling $\l \sim \mu^\nu$ 
 matches the gravitational dimension $ \delta_{1,3} = 1-\nu$, obtained from 
  $\Delta_{1,3}$ by the KPZ scaling relation $\Delta = {\delta(\delta+\nu)\over 1+\nu}$
\cite{KPZ}.  This is consistent with the conjecture that for  finite $\l /\mu^\nu$ the 
theory  is described by a perturbation of the critical point  $\l=0$ by the
Liouville-dressed thermal operator $\Phi_{1,3}$. 
An analysis of eq. (\ref{STREQ}) in the framework of Liouville gravity
was performed  by Al. Zamolodchikov \cite{Z3pt}. 
After integration one can reconstruct  the expansion
of the partition function $\CF$ in $\l$, which generates the $n$-point
correlation functions of the thermal operator. 
The result for $n\le 4$ matches with the
calculations recently published by A. Belavin and Al.  Zamolodchikov
\cite{PBZY} as well as with the formula conjectured in \cite{KP} on
the basis of the ground ring identities.    We leave
the analysis of the case of a strong perturbation from the perspective 
of the worldsheet CFT to a future publication \cite{KZ}.

\section{Solution of the $O(n)$ matrix model in the planar limit}

   \subsection{The $O(n)$ matrix model}

The partition function (\ref{lopgs}) of the $O(n)$ model on random
triangulations is generated as the perturbative (t'Hooft) expansion of
the free energy of a model of $n+1$ hermitian $N\times N$ matrix
variables $X$ and $\vec Y=\{Y_1, ..., Y_n\}$ \cite{KostovFY}:
 \eqn\partfXY{ \e^{\CF} \sim \! \int  \! dX\!
     \prod _{a=1}^n d { Y} _a\
    e^{ N\k^{-2} \({-{1\over 2}  \tr X^2 -{1\over 2} \sum _{a=1}^n {Y} _a^2
     +{1 \over 3}   X^3 + {1\over 2 T}    \sum_{a=1}^n X { Y} _a^2}\)}.
     \label{partfXY}
     }
 We are interested  in the planar limit $N\to\infty$, where only 
 genus zero planar Feynman graphs survive. Each such  Feynman  
 grah realizes a loop configuration in the sum (\ref{lopgs}). 
 
  Integrating out the $Y$-variables and shifting $X\to X+\hf T$
one obtain one-matrix integral  of the form
   \eqn\partf{
     \e^{ \CF} = \int  dX\   
     e^{-\b\, \tr V(X) }
    \ \left| \Det(1\otimes X + X\otimes 1)\right|^{-n/2}
   \label{partf}
     }
 where $\b = N/\k^2$ and the coefficients of the cubic potential
     \eqn\npotn{
     V(X)  = t_1X +  t_2 X^2 +  t_3 X^3 
   \label{npotn}
   } 
are expressed in terms of  the temperature $T$ as
\eqn\betattt{ 
    t_1=    \frac{T(2-T)}{4}     ,\  \
     \ t_2 =   \frac{1-T}{2}      , \  \
     t_3=-   \frac{1}{3} .
     \label{betattt} }
It will be convenient to take the large $N$ limit by sending $\b$ to
infinity keeping the ratio $\k^2 = \b/N$ finite.

After diagonalization $gXg^{-1}= \{x_1,...,x_N\} , \ \ g\in SU(N)$,
the matrix integral (\ref{partf}) can be reformulated as a
two-dimensional Coulomb gas of $N$ charges constrained at the real
axis:
\eqn\CoulombZ{
 e^{\CF} \sim 
 \int _{{\IR}}  
 \prod_{i=1}^{N}  dx_{i}
\ e^{-\b  V(x_{i})}  \  \prod_{i<j}^{N}
(x_{i}-x_{j})^2
 \prod_{i,j=1}^{N} 
  (x_{  i}+x_{ j})^{-{n\over 2}}.
\label{CoulombZ}
}
In the limit $N\to\infty$ the integral is saturated by the saddle
point described by the classical spectral density $\rho(x)$ supported
by the interval $[-a, -b]$ on the negative semi-axis, where $0<b<a$.
We will normalize the spectral density as
\eqn\densty{
  \int _{-a}^{-b}  dx\,  \rho(x) = \frac{N}{\b} = \k^2.
 }
The spectral density is determined from the saddle point equation 
\eqn\sdptCG{
V'(x) = 2\CP \int \limits_{-a}^{-b} dy {\rho(y)\over x-y} -n
\int \limits_{-a}^{-b} dy {\rho(y)\over x+y}, \qquad x\in[-a,-b].
\label{sdptCG}
}
The edges of the eigenvalue distribution are functions of the external
potential, {\it i.e.} of the temperature $T$ and the cosmological
constant $\k$.

The  disk partition function with boundary length $\ell$ is defined
as
\eqn\diskpf{
\tilde \Phi(\ell) =     \frac{1}{\b}\,  \tr e^{\ell X}
=\frac{1}{\b}   \int_{-a}^{-b} \, e^{\l\ell} \rho(\l)d\ell.
\label{diskpf}
}
The position of the right endpoint, $-b$, determines the 
large $\ell$ asymptotics of the disk partition function, $\Phi(\ell)\sim e^{- b \ell}$,
 and  has the statistical meaning of  boundary entropy for unit length. 
 The boundary entropy is negative because shifting the matrix variable
 $X$ we effectively performed a subtraction.

  Knowing the position of the edge $b=b(T,\k)$, the
equations of the two branches of the critical line 
 mentioned in the Introduction 
are obtained as follows
\cite{KostovFY}.  The  branch  $ \k = \k_{II}(T)$ is determined by the
condition that the right endpoint reaches the origin:
\eqn\Tkapa{
 b=0   \ \ \Rightarrow \ \   \k _c= \k_{II}(T).
 \label{Tkapa}
 }
When $b>0$ the loops are not critical, the critical behavior
is that of pure gravity $(c_{\rm matter }=0)$ and the position $\p_\k
b\sim \p_\k^2 \CF $ must behave as $(\k_c-\k)^{-1/2}$.
The  branch  $\k_c =\k_I(T)$ is determined by the condition  
that the position of the endpoint $b$ develops a singularity in $\k$:
\eqn\kapat{
 {\p b\over \p \k}\Big|_{T} \to\infty \ \ \Rightarrow \ \ \k_c = \k_I(T).
}
Finally, the   critical point $ \{T,\k\} =\{ T^*, \k^*\}$ is the common
 endpoint of the 
two critical lines: $   k^* = \k_I(T^*)= \k_{II}(T^*) $.

\subsection{Functional equation for the loop  field}

The saddle point equation can be solved directly,  see {\it e.g.} 
\cite{GaudinVX} for the solution in the special case $b=0$.
It is however more advantageous to reformulate the 
problem in terms of the collective field,
or {\it loop operator},
 \eqn\defPh{
  \Phi (x) =  -   \frac{1}{\b}   \sum _{i=1}^N  \log(x-x_i) ,
\label{defPh}
}
which is the Laplace transform of the disk partition function 
(\ref{diskpf}).
The geometrical meaning of the operator $\Phi$ is that it creates a
boundary with, in  general complex, boundary cosmological 
constant  $\mu_B =x$.   The  vacuum
 expectation value $\<\Phi(x)\>$ is equal to  the disk partition function,  
 the connected  correlator  $\<\Phi(x)\Phi(x')\>_c$ gives the partition 
 function on a cylinder,  etc.
 
The saddle point equation (\ref{sdptCG}) can be reformulated   
as a boundary condition for  the  current  
   \eqn\defJ{
   J(x) =   - \frac{2 V'(x) + nV'(-x)}{ 4-n^2 }
   - \p_x \Phi(x),\quad
  x\in\IC,
\label{defJ}
}
which is analytic on the complex plane cut along the interval $[-a,-b]$.
Namely, the values of $J(x)$ on both sides of the cut are related by
     \eqn\homfe{
     J(x+i0)+J(x-i0) +n J(-x)=0  \qquad x\in[-a,-b].
    \label{homfe}
     }
The function $J(x)$ is completely determined by the  boundary condition
 (\ref{homfe}) and the first four coefficients  of its  Laurent  expansion at infinity, 
\eqal\Jinf{ J(x) =   \sum_{n= -\infty}^3 J_{n}\,  x^{n-1}=
J_3\, x^2 + J_2 \, x + J_1\, + J_0\, x^{-1} + [J(x)]_{<0},
    \label{Jinf}
     }
which  are  expressed in terms of  $T$  and $\k$,
 \eqal\Jnnn{  J_0 = \k^2, \ \ 
  J_1
  =   -  \frac{ T(2-T)}{4 (2-n)}  , \ \ 
   J_2
   =  \frac{T-1}{2+n},\ \ 
    J_3
    = \frac{1}{2-n}.
 \label{Jnnn}
 }

We will split, as in \cite{EynardCN}, the function $J(x)$ in two
`chiral' pieces,
     \eqn\JJJ{
     J(x)=J_+(x)+J_-(x),
     }
    satisfying  simpler  boundary conditions along the cut:
     \eqal\fucpm{ \begin{matrix}
     J_+(-x\pm i0) &= -e^{\pm i\pi\nu} \ J_+(x\pm i0)\cr
     J_-(-x\pm i0) &= -e^{\mp i\pi\nu} \ J_-(x\pm i 0) 
     \end{matrix}
     \qquad (b<x<a).
     \label{fucpm}
     }
    The phase $\nu$ is defined by (\ref{nnu}).
The boundary condition (\ref{homfe}), respectively (\ref{fucpm}),
implies a quadratic functional identity for the current
\cite{KostovCG}
  \eqn\qfue{\hskip -0.14cm
   \[J^2\! (x)+J^2\! (-x) +nJ(x)J(-x)\]_{<0} \! \equiv  (2-n^2)  [J_+(x)J_-(x)]_{<0}=0,
     \label{qfue}
     }
where $[\ \ ]_{<0}$ denotes the negative piece of  Laurent expansion. 
Therefore $J_+(x)J_-(x) $ is an even polynomial of 4 degree
 whose  coefficients   are  known functions of $\k$ and $T$.
 In order to avoid heavy 
expressions  we  rescale $x$ and $b$ as
 \eqn\hhat{
 \hat x={x/ a},
 \qquad  \hat b =  {b/ 4a} 
 }
and use  directly the  Laurent expansion of the chiral components 
$J_{\pm}$:
\eqn\exJpm{
 J_\pm(x) = e^{\mp i \pi \nu/2}\( \sum_k \hat J_{2k} \hat x^{2k-1}
 \pm i \sum_{k } \hat J_{2k+1} \,  \hat x^{2k}\).
 \label{exJpm}
}
The new expansion coefficients are related to the old ones by
\eqal\hatJs{ 
     \hat J_{2k} \ &=\frac{a^{2k-1} }{\sqrt{2-n}}J_{2k} ,\qquad 
\hat J_{2k+1} = \frac{a^{2k}}{ \sqrt{2+n} }J_{2k+1}.
\label{hatJs}
}
 Then  (\ref{qfue}) implies the   functional equation
 \eqn\Ppmi{
 J_+(x) J_-(x) =  A +  B \hat x^2 + C\hat x^4
 \label{Ppmi}
 }
where the  $A,B$ and $C$  are expressed in terms of 
$ \hat J_3,  \hat J_2,  \hat J_0$ and the first moment of the eigenvalue density
$  \hat  W_1\equiv   \hat  J_{-1}$:
\eqal\eqnsym{
   A = {2\hat W_1 \hat J_3 +\hat J_1^2 } + 2\hat J_0 \hat J_2 ,\quad
   B=2  {\hat J_1\hat J_3}  + {\hat J_2^2 } ,
   \quad
   C=  \hat J_3^2  
\label{eqnsym}
}

\subsection{A criterium for criticality}

 The functional equation (\ref{Ppmi}) leads to  an algebraic equation for $J(0)$:
           \eqn\crtpt{
J^2(0)\equiv (2-n) \ J_+(0)J_-(0) =(2-n) \ A\ \Rightarrow \ J(0) =   \sqrt{(2-n)A}
      .
      }
The value $J(0)$ is analytic function of the couplings $\k^2$ and $T$ 
in the domain where the loop gas partition function is convergent. 
The critical lines are therefore given by the   border of the domain 
of analyticity of $J(0)$.   The latter is given by the condition  $A=0$, or
 \eqn\kappac{
   2   \hat W_1 \hat  J_3 + \hat J_1^2+
      2  \hat  J_0\hat J_2
     =0 .
     }

\subsection{Solution along the critical line $\k =\k_{II}(T)$}

 \def\hx{ \hat x } 

\noindent
The Riemann surface of $J(x)$ consists 
of infinitely many sheets, except for the case when $\nu$ is a rational number.  
The first, physical, sheet has one cut
$[-a, -b]$ while  all the other sheets have two cuts $[-a, -b]$ and
$[b,a]$ (Fig.  2).  If we  find  a global
parametrization of the Riemann surface that resolves all branch
points, then the boundary conditions (\ref{fucpm}) will  become  quasi-periodicity
conditions and can be solved (in our case) in terms of theta functions.

  \epsfxsize=200pt
 \vskip 20pt
 \hskip 100pt
 \epsfbox{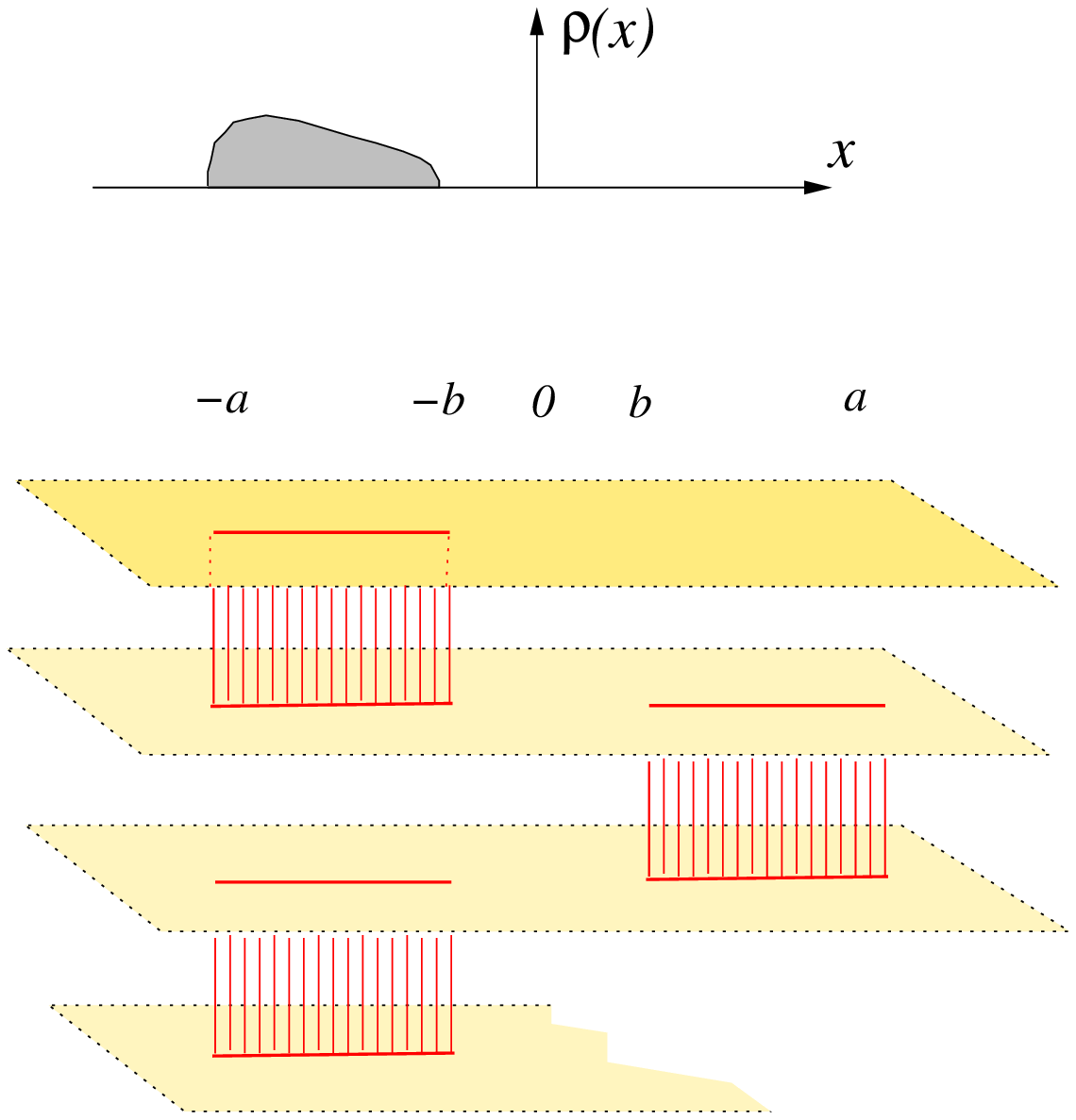  }
 \vskip 5pt
 
 \centerline{Fig. 2:   The Riemann surface of $J(x)$    }
  
  \vskip  10pt

%

To illustrate the method, let us  first reproduce the solution
along the critical line $T=T_c(\k)$,  found originally in 
 \cite{GaudinVX} by applying the Wiener-Hopf method to 
 the integral equation (\ref{sdptCG}). Along this critical line $b=0$ and
  the infinity of simple  ramification  points at $x=\pm b$ merge into 
 a single ramification point of infinite order at $x=0$.
 The Riemann surface  is  then globally parametrized by a hyperbolic map
\eqn\trigx{
 \hat x(s)= {1\over \cosh s}
}
where the points with parameters $s$ and $-s$ must be identified.  The
boundary conditions (\ref{fucpm}) become a quasi-periodicity conditions
in the complex $s$-plane:
\eqal\Jose{ 
      J_{+}(s\pm i \pi  ) &=- e^{\pm i\pi\nu}\ J_+(s),\cr
      J_-(s\pm i\pi ) &= -e^{\mp i\pi\nu}\ J_-(s).
      \label{Jose}}
The unique solution of (\ref{Jose})  with the asymptotics  (\ref{exJpm})
is
\eqal\Jpmc{  
{J_\pm(s) } =  {d_1  } { e^{ \pm \nu s}\over \cosh s}
+ {d_2  } { e^{\pm( \nu-1) s}\over \cosh ^2 s}.
}
 with
 \eqn\dadb{
  d_2= \hat J_3 ,\ \ 
 d_1=  \hat J_2- (1-\nu)\hat J_3.
 \label{dadb}
 }
  One can check that the 
 quadratic equation (\ref{Ppmi}) is satisfied with 
 $$A= 0, \ \ \ B=  d_1(2d_2+d_1), \ \ \ C= d_2^2.$$
Comparing with the asymptotics  (\ref{exJpm}) we find two more relations
 \eqal\Jndn{
 \hat J_1 &=&     \nu \hat J_2 -\frac{1-\nu^2}{2} \hat J_3
 ,\ \ \ \ 
 \hat  J_0=
   \frac{\nu^2}{2} \hat J_2-  \frac{1-\nu^2}{3}\hat J_3,}
 which  determine  the left branchpoint  $a=a(\k)$   
 and the critical  line   $ \k _c = \k_{II}(T)$ that describes the dense phase. 
 The   line $ \k _c = \k_{II}(T)$  starts at the critical point  $T^*$ where $d_1=0$.

In a similar fashion we  calculate the $J_0$-derivative of $J(z)$,
which we rescale  for convenience as 
\eqn\defhatJ{
 \hat \p_0J(x)
\equiv  {a \sqrt{2-n}} \ \p_0J(x) .
}
 The meromorphic functions $\hat  \p_0J_\pm(x)$ satisfy the same
boundary condition (\ref{Jone}) but have less singular behavior at
infinity: 
     \eqn\dNJpm{
      \hat \p_0J _\pm (x)=   \pm i  {e^{\mp i \pi \nu/2} }
      \( 
    \frac{1}{ \hat x}\mp i    \frac{\p_0 \hat  W_1}{ \hat  x^2}+ \dots \)
    , \qquad x\to\infty.
    \label{dNJpm}
    }
   The only solution of  (\ref{Jone}) with such asymptotics  is  
\eqal\Jhis{ \hat \p_0J_\pm= {\pm i } 
e^{\mp  \nu s }  \coth s
.
}
We finally obtain for the current (\ref{defJ}) and  its derivative  
\eqal\resolvc{
 J\ \  &=& 
2d_1 \frac{\cosh \nu s}{  \cosh s} + 2 d_2 \frac{\cosh (1-\nu)s}{  \cosh^2 s}
\cr 
     \hat  \p_0J&= & 
      \sinh \nu s \, \coth s
     .
\label{resolvc}
}
The scaling behavior  for  $|\hat x|\ll1$  is  
\eqal\scalJ{
  &&{J(x) }\ \ \
    \sim     ( \mu_B )^{1+\nu} 
       -    \l \,   (\mu_B)^{1-\nu}   ,
       \quad  \cr
 && \p_0J(x) \sim   
  ( \mu_B)^\nu.
 \label{scalJcr}
 }       
where
\eqn\cplngs{ 
  \l 
 = \frac{1- \nu}{2} - \frac{\hat J_2}{2 \hat J_3}
 \sim T_*-T, 
 \quad  \mu_B = \hat x/2.
 \label{cplngs}
 }

    \subsection{The general solution}

A solution of the $O(n)$ model in the most general case is presented in
 \cite{ EynardNV, EynardZV}. However,  we found 
 easier to perform an independent computation rather than to use 
 the results of  \cite{ EynardNV, EynardZV}.  
    
For generic values of the couplings there exists a global
parametrisation in terms of Jacoby elliptic functions.  
The uniformization map $u\to x(u)$ is defined as
\eqn\xofu{
      x= {b\over \dn (u, k)},\qquad
       \qquad k'  \equiv \sqrt{1- k^2}= \frac{b}{a}
      \label{xofu}
      }
(our notations are those of Gradshtein and Ryzhik \cite{GRRZ}).

The map  has the symmetries $ x(u)=x(-u)=x(u+2K)=-x(u+2iK')$ and parametrizes
the physical sheet by the rectangle $[0,K]\times[-2iK',2iK']$ in the
$u$-plane.  The whole Riemann surface is parametrized by the orbifold
of the $u$-plane with respect to the symmetry $ u\to -u$.

The functions $J_\pm(u)$ must be periodic in $u\to u+2K$
\eqn\Jtwo{
      J_\pm(u+2K) = J_\pm(u)
      }
and, by the boundary condition (\ref{fucpm}), quasi-periodic in $u\to
u\pm 2iK'$:
         \eqal\Jone{ 
      J_{+}(u\pm 2iK') &=- e^{\pm i\pi\nu}\ J_+(u),\cr
      J_-(u\pm 2iK') &= -e^{\mp i\pi\nu}\ J_+(u).
      \label{Jone}}
The general solution of these equations is given in terms of the theta
function $Y_\o(u) $ defined as follows:
\eqn\defymu{
      Y_\o (u)= \frac{\Theta_1(0)}{ \Theta_1(\o  K)}
      \,   \frac{\Theta_1( u-\o  K)}{\Theta_1(u)}
       \qquad \Theta_1(u) = \theta_3\( \frac{\pi u}{ 2K}\).
      \label{defumu}
      }
The map (\ref{xofu}) is a particular case of this function, $x(u) = b
\, Y_1(u)$.  We will need the following properties of the function
$Y_\o(u) $: 

\vskip 4pt
 
 1) {\it Symmetries and (quasi) periodicity:}
  \eqal\pery{ 
    &  \yy(u)= {Y_{-\o } (-u)} =Y_{\o +2}(u)= \yy(u+2K)
     \cr
     &\cr
    &   \yy(u\pm 2iK')=e^{\pm i\pi\o } \yy(u)\hskip 2.2cm
            \label{pery} }

2) {\it Quadratic relations}
       \eqn\qfuc{
  Y_{ \o}(u) Y_{-\o}(u)   =   1+  \sd^2 (\nu K) \({\hat x^2- k'^2 }\) ,
    }
  \eqn\scrp{   Y_{1- \o}(u) Y_{\o}(u) 
  + Y_{-1+ \o}(u) Y_{-\o}(u) =  \frac{2}{ \k'} \ \hat x
  \label{scrp}
  }
       \eqn\YYdver{    Y_{1- \o}(u) Y_{\o}(u) 
  - Y_{-1+ \o}(u) Y_{-\o}(u) 
  =    \frac{1}{ k'} \frac{\sn\cn}{\dn}(\o K)\, {\p_u \hat x , }
   \label{qfuc}}

     \medskip

  3) {\it  Particular values:  }

 \vskip 10pt
\centerline{ \hbox{\qquad 
\vbox{\offinterlineskip
\hrule
\halign{&\vrule#&\strut\quad\hfil#\quad\cr 
height2pt&\omit&&\omit&&\omit
&&\omit&\cr
& $u $ \hfil && $x $ \hfil && $\yy $   \hfil & 
 \cr
height2pt &\omit &&\omit&&\omit&&
\omit&\cr
\noalign{\hrule}
&\omit&&\omit && \omit&\cr
& $0$ && $b $ && $  1 \ \ \ \qquad $&\cr
 & $  K$ && $ a $  && $    {\rm nd} (\o  K) \ \ \  $ 
 &\cr
         & $  K\pm iK' $ && $ \infty $  && $  \infty\qquad$ 
 &\cr
       & $  K\pm 2iK' $ && $ -a $  && $   e^{\pm i\pi\o }  {\rm nd} (\o  K)  $ 
 &\cr
  & $\pm 2iK'$ && $-b $ && $ e^{\pm i\pi\o }  \quad $ 
&\cr
 &\omit&&\omit&&\omit&&\omit&\cr}\hrule}
}
 }

\bigskip

    4)  {\it Asymptotics at $x\to\infty$:} 
       \eqal\yyx{
        \yy(x)  =  e^{ i\pi{ \o-1\over 2}} \, \sd(\o K)\
        \(  \hat  {x } - i   \frac{H'(\o K) }{H(\o K) }  +...\)
        \label{yyx}
        }    

One can easily check that the solution of (\ref{Jone}) in the case of
cubic potential is given by the linear combination\footnote{The
solution for a generic polynomial potential is obtained by replacing
the coefficients $D_1$ and $D_2$ with entire functions of $x^2$
\cite{EynardNV}.}
\eqal\Jpl{  J_\pm (u)&= D_1  Y_{\pm (1- \nu)} (u) 
+ D_2 Y_{\mp \nu}(u) \hat x(u).
\label{Jpl}
}
 Since $J_+(-u)=J_-(u)$, the
 function $J=J_++J_-$ is even function of $u$, hence a function on the
 orbifold.  The coefficients $D_1$ and $D_2$ and the  branch points   $a$
 and $b=k' a$   are obtained by from the asymptotics 
 at infinity given by (\ref{Jinf}).  To simplify the expressions, we will introduce the
 rescaled coefficients
\eqal\SnDn{ 
    d_1= D_1\frac{\cn(\nu K)}{ k'} , \qquad
    d_2= D_2\,   \sd(\nu K)  .
    \label{SnDn}
    }
Comparing the coefficients of the two leading  of the expansion at $x\to\infty$  
we find
\eqn\SoneStwo{
     d_1  =  \hat J_2-   \frac{H'}{H}(\nu K)  \hat J_3,
     \ \ \ \  d_2= \hat J_3.
     \label{SoneStwo}
     }
Note that (\ref{dadb}) = (\ref{SoneStwo}) $|_{k'=0}$.  We need two
more relations that determine the positions $a$ and $b$ of the branch
points.  Instead of expanding further around the point $x=\infty$, it
is simpler to use the fact that the solution (\ref{Jpl}) satisfy a
quadratic relation whose l.h.s. is identical to that of (\ref{Ppmi}).
This quadratic relation follows from (\ref{qfuc}) and (\ref{scrp}):
       \eqal\eqnsm{ 
   J_+J_-\! &=d_2^2 \hx^2 (\hx^2 \! + \cs^2(\nu K) ) + 2 d_1 d_2
   \frac{\dn}{\sn\cn}(\nu K)\, \hx^2  \cr
   &+ d_1^2(\hx^2\! +  k'^2 \sc^2(\nu K) ).\qquad \qquad 
 }
Taking $x=0$ we obtain  for the coefficient $A$
 \eqn\xzero{
 A\equiv   {2\hat W_1 \hat J_3 +\hat J_1^2 }+
         2{\hat J_0\hat J_2}
    =      k'^2   \sc^2 (\nu K)  d_1 ^2 .
         \label{xzero}
         }
The condition for criticality  $A=0$  is achieved either if $k'=0$ (dense phase), or
if $d_1=0$ (pure gravity), or if $k'=d_1=0$ (critical point, or dilute
phase).  
To evaluate the boundary entropy $M\sim b(\l,\mu)$ we will proceed as follows. 
 We will  first  calculate independently $J$ and $\p_\mu J$, then take the continuum limit and compare the second quantity with the derivative of the first in $\mu$. This will give an equation for the function $b(\l,\mu)$.

\bigskip

In a similar way we  evaluate  the  derivative  $ \hat \p J_\pm (x)$
as the unique solution of (\ref{Jone}) and  
(\ref{dNJpm}).    It  is given by  
    \eqn\pJ{\hat  \p_0J _\pm=  
  \pm i \ds(\nu K)      \,  {Y_{\mp \nu}(u)\over  \p_u \hat x}.
 \label{pJ}
  }
 where
 $$  \p_u \hat x =i \sqrt{(\hat x^2 -1)( \hat x^2 - k'^2)}.$$
The solution (\ref{pJ})  leads to a simple formula for $ \hat\p_0\p_u \Phi  $:
\eqn\dmPhi{  
      \hat\p_0\p_u \Phi  =
-  \ds(\nu K) \, (Y_{ \nu}- Y_{-\nu})
  .
 \label{dmPhi}   }
Finally,  comparing the subleading terms  of the expansions (\ref{dNJpm})
and (\ref{yyx}), we find the derivative of  $W_1 = {1\over \b} \langle \tr M\rangle$,
    \eqn\dNWone{
 - \hat \p_0 \hat W_1=   
      { H'(\nu K)\over H(\nu K)} 
    \label{dNWone}
       ,}
 whose singular part  gives, up to a normalization, the 
susceptibility $\chi\sim-  \hat \p_0^2\CF$.

\subsection{Continuum limit}

 The continuum  limit  $a\to\infty$ is  achieved when both
  $ \hat x = x/a$ and $\hat b = b/ 4a
 = k'/4$  are small. In this limit  the uniformization map (\ref{xofu}) 
 degenerates to 
   \eqal\xuu{
   x(u)  =  b \cosh u 
   }
 and the  theta function (\ref{defumu}) is approximated by
   \eqal\Ycl{
    Y_{\pm \o}(u) =
  e^{\pm \o u}
. 
  }  
  The two coefficients of  the   solution  (\ref{Jpl})   are now given by
 \eqn\DDdd{
  D_1= 2\hat b^{1-\nu} d_1, \ \ \ \ 
 D_2= 2 \hat b^{\nu} d_2 
 }
and for the current $J=J_++J_-$ we find 
   \eqal\scaJj{
 J&=4 \hat J_3
 \(-\l \,   \hat b^{1-\nu} \cosh (1-\nu)u
 +   \hat b^{1+\nu}     \cosh(1+\nu) u \)
\label{scalJ} 
}
with  the  coupling $\l$  defined  earlier in  (\ref{cplngs}).
  On the other hand, from (\ref{pJ}) we find for the derivative $\hat \p_0 J$
  \eqn\doJ{
  \hat \p_0 J=   \hat b^{\nu -1}\, {\sinh \nu u\over\sinh u} .
  \label{doJ}
  }

 Now let us  compare (\ref{doJ}) with the derivative of (\ref{scalJ}):
\eqal\bzbz{
\hat \p_0 J|_x
&=&    {\hat \p_0 \hat   b } \,  \( \frac{\p}{ \p \hat b}
 - \frac{\p_u }{ b  \tanh   u} \)\cdot J
\cr 
&=& 4 \hat J_3\  {\hat \p_0  \hat b } \,  \( - \l  (1-\nu)  \hat b^{-\nu}  
- (1+\nu)  \hat b^\nu\)
 {\sinh \nu  u\over \sinh u} .
}
  The two expressions coincide under the condition
  \eqn\compcon{
  \hat J_0 -\hat J_0^c=- 2 \hat J_3 \( \l \hat b^{2-2\nu} + (1+\nu) \hat b^2\),
\label{compcon}  }
   where the $T$-dependent integration constant $\hat J^c_0$ corresponds to
 the value $\k _c = \k_{II}(T)$.

 Eq. (\ref{compcon}) is identical to the transcendental equation 
   (\ref{bentr}) 
    with   the cosmological  constant $\mu$ and the   boundary 
 entropy $-M$  being normalized as
 \eqal\mulaM{
 \mu = \frac{\hat J_0^c- \hat J_0}{2  \hat J_3} , \qquad M = \hat b.
 }
 The susceptibility $\chi= - \p_\mu^2\CF$ is  given, up to a constant factor,
  by the leading 
 singular term in the small $k'$ expansion 
   of (\ref{dNWone}):
  \eqn\dWcl{
   - \hat \p_0 \hat W_1=   
           1-\nu + 2 \hat b^{2\nu}+...
      }
 Since we have not yet normalized  the string interaction constant $g_s \sim \b$,
 we have the freedom to normalize the free energy. We do it so that
 $ \chi = M^{2\nu}$.

\section*{Acknowledgments}

The author is obliged to  Al. Zamolodchikov for many enlightening discussions. 
This work has been partially supported by the  European Union
 through the FP6 Marie Curie RTN { ENIGMA} (contract MRTN-CT-2004-5652), 
 ENRAGE (contract MRTN-CT-2004-005616) and  the ANR program
  "GIMP" (contract ANR-05-BLAN-0029-01).

\end{document}